\documentclass[12pt]{article}

\usepackage{epsfig,layout}
\begin{document}

\title{Universal Codes as a Basis for  Time Series Testing  }

\author{Boris Ryabko
      and  Jaakko Astola
\thanks{Research  was supported  by the joint
project grant "Efficient randomness testing of random and
pseudorandom number generators" of  Royal Society, UK (grant ref:
15995) and Russian Foundation for Basic Research (grant no.
     03-01-00495.).}
}

\date{}
\maketitle

\begin{abstract}
We  suggest a new approach to  hypothesis testing for ergodic and
stationary processes. In contrast to standard methods, the
suggested approach gives a possibility to make tests, based on any
lossless data compression method even if the distribution law of
the codeword lengths is not known. We apply this approach to the
following four problems: goodness-of-fit testing (or identity
testing), testing for independence,  testing of serial
independence
 and  homogeneity testing  and suggest nonparametric statistical
 tests for these problems. It is important to note that practically used
so-called archivers can be used for suggested testing.

\end{abstract}


\textbf{keywords} universal coding, data compression,  hypothesis
testing, nonparametric testing, Shannon entropy, stationary and
ergodic source.


\section{Introduction}

Since Claude Shannon  published his famous paper "A mathematical
theory of communication" \cite{S}, the ideas and results of
Information Theory have begun to play an important role in
cryptography \cite{Ma, S2}, mathematical statistics \cite{Ba, CS,
Ku, Ri2}, ergodic theory \cite{Ba,Billingsley, Shi} and many other
fields \cite{V1,V2,RR} which are far from telecommunication. The
   theory of universal coding, which is a part of
Information Theory, also has been efficiently applied to many
fields since its discovery in \cite{Fi,K}. Thus, application of
results of universal coding, initiated in \cite{ Ry1}, created a
new approach to prediction \cite{Ki, Mo, No}.

In this paper we suggest a new approach to hypothesis testing,
which is based on ideas of universal coding. We would like to
emphasize that,  on the one hand, the problem of hypothesis
testing is considered in the framework of classical mathematical
statistics and, on the other hand, everyday methods of data
compression (or archivers) can be used as a tool for testing. It
is important to note that the modern archivers  are based on deep
theoretical results of the source coding theory (see, for ex.,
\cite{e, K-Y,Kr,Ri,Sa}) and have shown their high efficiency  in
practice as compressors of texts, DNA sequences and many other
types of real data. In fact, universal codes and
 archivers can find  latent regularities of many kinds, that is
why they  look like a promising tool for hypothesis testing.

\subsection{The main idea of the suggested approach}

Let us describe the main idea of the suggested approach using
one particular  problem of hypothesis testing which is
conceptually simple and yet is important for practise. Namely,
we consider a null hypothesis
$H_0$ that a given bit sequence $x_1 ... x_t$ is generated by
a Bernoulli source with equal probabilities of 0 and 1 and the
alternative hypothesis
$H_1$ that the sequence is generated by stationary and ergodic
source, which differs from the source under $H_0$. This problem is
considered  in \cite{RM} and is a particular case of the
goodness-of-fit testing
 ( or identity testing) described below, that is why we give an informal
solution only. Let $\varphi$ be a universal code, $\varphi(x_1 ...
x_t)$ be the encoded sequence, $l_\varphi(x_1 ... x_t)$ be the
 length of the word $\varphi(x_1 ... x_t)$
  and $\alpha$ be the required level of significance.
Intuition suggests that the sequence cannot be compressed if
$H_0$ is true, and vice versa, if the sequence can be compressed
$H_0$ should be rejected. The corresponding formal test is as
follows: if $ ( t - l_\varphi(x_1 ... x_t) )
> \log (1/ \alpha), $ then  $H_0$ should be rejected.
(Here and below $\log \equiv \log_2.$) It
will be proven below that the
 Type I error of this test is equal to or less than $\alpha$ for
 any (uniquely decodable) code $\varphi$, whereas the
 Type II error goes to 0 for any universal code $\varphi,$ when
 the sequence length $t$  grows.

Let us look at the described test in more details. It is well
known that
 the average codeword length  of any code is not less than the
 sequence length $t$, if $H_0$ is true. Hence, if we define the
 codeword length of the best code as $l_{H_0}(x_1 ... x_t)$, we can see that
$l_{H_0}(x_1 ... x_t) = t$. Now the scheme of the  suggested test
can be described as follows: \emph{If $l_{H_0}(x_1 ... x_t) -
l_\varphi(x_1 ... x_t) \leq \log (1/ \alpha)$ then $H_0,$
otherwise $H_1$.} We will apply this scheme to all considered
statistical problems, sometimes replacing the length
$l_{H_0}(x_1 ... x_t)$ with its lower bound (as a rule, such a
lower bound will be based on   so-called empirical Shannon
entropy).

\subsection{Description of considered problems}

We consider a stationary and ergodic source (or process), which
generates elements from some set (or alphabet) $A$ (which can be
either finite  or infinite) and four problems of statistical
testing.

The first problem is the goodness-of-fit testing (or identity
testing), which is described as follows: a hypotheses $H_0^{id}$
is that the source has a particular distribution $\pi$
and the alternative hypothesis $H_1^{id}$ is that the sequence
is generated by a stationary and ergodic source which differs
from the source under $H_0^{id}$. One particular case, in
which the source alphabet
$A$ equals $\{ 0, 1 \} $ and the main hypothesis $H_0^{id}$ is that a bit
sequence is generated by the Bernoulli source with equal
probabilities of 0's and 1's, was mentioned in Introduction.

 The second problem is a generalization of the problem of
nonparametric testing for serial independence of time series. More
precisely, we consider the two following hypotheses: $H_0^{SI}$ is
that the source is Markovian of order   not larger than $m,\: (m
\geq 0),$ and the alternative hypothesis $H_1^{SI}$ is that the
sequence is generated by a stationary and ergodic source which
differs from the source under $H_0^{SI}$. In particular, if $m=0,$
this is the problem of testing for independence of time series.

The third problem is the  independence testing. In this case it
is assumed that the source is Markovian, whose order  is not
larger than $m,\: (m \geq 0),$ and the source alphabet can be
presented as a product of $d, d \geq 2, $ alphabets $A_1, A_2,
\ldots, A_d$ (i.e. $A = \prod_{i=1}^d A_i$). The main hypothesis
$H_0^{ind}$ is that $ p(x_{m+1} = (a_{i_1}, \ldots, a_{i_d}) |\,
x_1 ... x_m ) $ $ = $ $ \prod_{j=1}^d  p(x_{m+1}^{(j)} =
a_{i_j}|\, x_1 ... x_m ) $ for each $(a_{i_1}, \ldots, a_{i_d})
\in \prod_{i=1}^d A_i,$ where $x_{m+1}= $ $ (x_{m+1}^{(1)}, $ $
... ,$ $ x_{m+1}^{(d)}).$ The alternative hypothesis $H_1^{ind}$
is that the sequence is generated by a Markovian source of order
  not larger than $m,\: (m \geq 0),$ which
differs from the source under $H_0^{ind}.$

In all three cases the testing should be based either on one
sample $x_1 \ldots x_t$
 or on a several ($l$) independent samples $x^1=
x^1_1 \ldots x^1_{t_1}, $ $\ldots$ $x^l= x^l_1 \ldots
x^l_{t_l} $ generated by the source.
\footnote { For a case of
one sample and a finite alphabet $A$ some of these  problems
were considered by the authors in \cite{RA} and reports
submitted to conferences.}

The fourth problem is the homogeneity testing. There are $r$
samples $x_1^1 \ldots x_{t_1}^1,$ $x_1^2 \ldots x_{t_2}^2, ... ,$
$x_1^r \ldots x_{t_r}^r $ and it is assumed that they are
generated by
 Markovian sources, whose orders  are not larger than $m,\: (m
\geq 0).$ The main hypothesis $H_0^{hom}$ is that all samples are
generated by one source,  whereas the alternative hypothesis
$H_1^{hom}$ is that at least two samples are generated by
different sources.

All four problems are well known in mathematical statistics and
there is an extensive literature dealing with their nonparametric
 testing, see for review, for example, \cite{GKR,KS}.

\subsection{Main results} We suggest statistical tests for all
problems such that the Type I error is less than or equal to a
given $\alpha$ and the Type II error goes to zero, when the
sample size grows. However, there are some additional
restrictions mainly concerned with the case of infinite source
alphabet. For this case all test are described for memoryless
(or i.i.d.) sources only. It is important to note that the
suggested tests are based on universal codes (and closely
connected universal predictors), but the Type I error is less
than or equal to a given $\alpha$ for any code and, in
particular, it is true for practically used methods of data
compressions (or archivers), that is why they can be used as a
basis for the tests.
\subsection{Outline of the paper}
  The next section   contains some
necessary facts and   definitions. The sections three and four are
devoted to description of the tests for the cases  where alphabets
are finite  and infinite, respectively. Some experimental results
and simulation studies are given in the section 5.

 We  give a description of one particular universal code in
Appendix 1,
 because universal codes play a key role in this paper, but
 information about them is spread between numerous papers and
 they are not widely presented in statistical literature (in spite of the fact that
 universal codes have found
 different applications to some classical problems of mathematical
 statistics, see, for ex.,\cite{CS}). Besides, the universal code described in Appendix
 1 is used for simulation study of  serial independence testing in
 the part 5. (On the other hand, this paper focuses on
 hypothesis testing, that is why description of the universal
 codes and  ideas  behind them are put in the appendix.)

 The conclusion is intended to clarify the connection of the suggested approach
 and briefly describe some possible generalizations of the described tests.
  All proofs are given in Appendix 2.

\section{Definitions and Auxiliary Results }

\subsection{Stochastic  processes and the Shannon entropy}
Now we briefly describe stochastic processes (or sources of
information). Consider an alphabet $A,$ which can be either
finite or infinite,
 and denote by $A^t$ and $A^*$ the set of
all words of length $t$ over $A$ and the set of all finite
words over $A$ correspondingly ($A^* = \bigcup_{i=1}^\infty
A^i$). By
$M_\infty(A)$ we denote the set of all stationary and ergodic
sources, which generate letters from $A;$ see for definition, for
ex., \cite{Billingsley,Ga} and let $M_0(A) \subset M_\infty(A)$ be
the set of all i.i.d. processes.
 Let $M_m(A) \subset M_\infty(A)$ be the set of Markov
sources of order (or with memory, or connectivity) not larger than
$m, \, m \geq 0. $ In the case of a finite alphabet $A$ Markov
processes will play a key role in this paper, that is why we
give  a formal
 definition. By definition $\mu \in M_m(A)$ if
\begin{equation}\label{ma}
\mu (x_{t+1} = a_{i_1} |\, x_{t} = a_{i_2}, x_{t-1} = a_{i_3},
\dots, x_{t-m+1} = a_{i_{m+1}},\dots ) \end{equation}$$
 = \mu (x_{t+1} =
a_{i_1}|\, x_{t}  = a_{i_2}, x_{t-1} = a_{i_3},\dots x_{t-m+1}
= a_{i_{m+1}})
$$
 for all $t \geq m $ and
$a_{i_1}, a_{i_2}, \ldots \, \in A.$ Let
$M^*(A) = \bigcup_{i=0}^\infty M_i(A)$ be the set of all
finite-order sources.

Let $\tau$ be a stationary and ergodic source generating letters
from a finite alphabet $A$. The $m-$ order (conditional) Shannon
entropy and the limit Shannon entropy are  defined as follows:
\begin{equation}\label{moe} h_m(\tau) =
\sum_{v \in A^m} \tau(v) \sum_{a \in A} \tau(a|\,v) \log
\tau(a|\,v),\qquad
 \: h_\infty(\tau) = \lim_{m \rightarrow \infty} h_m(\tau). \end{equation}
 It is also known that for any $m$
\begin{equation}\label{hlim}
h_\infty(\tau) \leq h_m(\tau) \:, \end{equation} see
\cite{Billingsley, Ga}. The well known
Shannon-MacMillan-Breiman theorem states that
\begin{equation}\label{smb}
\lim_{t\rightarrow\infty} -  \log \tau(x_1 \ldots x_t) /t =
h_\infty(\tau)
\end{equation} with probability 1, see \cite{Billingsley, Ga}.

  Let $v= v_1 ... v_k$  and $x = x_1 x_2 \ldots x_t$ be words from
$A^* .$ Denote the rate of a word $v$ occurring in
the sequence $x = x_1 x_2 \ldots x_k$ , $x_2x_3 \ldots x_{k+1}$,
$x_3x_4 \ldots x_{k+2}$, $ \ldots $, $x_{t-k+1} \ldots x_t$ as
$\nu_x(v)$. For example, if $x = 000100$ and $v= 00, $ then $
\nu_x(00) = 3$. For any $0  \leq k < t$ the  empirical Shannon
entropy of order $k$ is defined as follows:
\begin{equation}\label{He}
h^*_{ k}( x) = -  \sum_{v \in A^k} \frac{\bar{\nu}_x(v)}{(t-k)}
\sum_{a \in A} \frac{\nu_x(va)}{ \bar{\nu}_x(v)} \log
\frac{\nu_x(va)}{ \bar{\nu}_x(v)}\, ,
\end{equation}

 where $x = x_1 \ldots x_t,$ $ \bar{\nu}_x(v  )= \sum_{a \in A} \nu_x(v a ). $
In particular, if $k=0$, we obtain $ h^*_{ 0}( x) = - t^{-1}
\sum_{a \in A} \nu_x(a)  \log (\nu_x(a) / t )\, . $

We extend these definitions to a case where a sample is presented
as several (independent) sequences $x^1= x_1^1 \ldots x_{t_1}^1,$
$x^2 = x_1^2 \ldots x_{t_2}^2, ... ,$ $x^r = x_1^r \ldots
x_{t_r}^r $ generated by a source. (The point is that we cannot
simply combine all samples into one, if the source  is not i.i.d.)
We denote this sample by $x^1 \diamond x^2 \diamond \ldots
\diamond x^r$ and define $t = \sum_{i=1}^r t_i,\,$
 $\nu_{x^1\diamond
x^2\diamond ... \diamond x^r } (v) = \sum_{i=1}^r \nu_{x^i}(v) .$
For example, if $x^1 = 0010, x^2 = 011,$ then $\nu_{x^1\diamond
x^2}(00)= 1.$ Analogously to (\ref{He}),
\begin{equation}\label{He1}
h^*_{ k}( x^1\diamond x^2\diamond ... \diamond x^r ) = -  \sum_{v
\in A^k} \frac{\bar{\nu}_{x^1\diamond
 ... \diamond x^r }(v)}{(t-k r)} \sum_{a \in A}
\frac{\nu_{x^1\diamond  ... \diamond x^r }(va)}{
\bar{\nu}_{x^1\diamond  ... \diamond x^r }(v)} \log
\frac{\nu_{x^1\diamond  ... \diamond x^r }(va)}{
\bar{\nu}_{x^1\diamond  ... \diamond x^r }(v)}\, ,\end{equation}
where  $ \bar{\nu}_{x^1\diamond  ... \diamond x^r }(v  )= \sum_{a
\in A} \nu_{x^1\diamond  ... \diamond x^r }(v a ). $

For any sequence of words  $x^1= x_1^1 \ldots x_{t_1}^1,$ $x^2 =
x_1^2 \ldots x_{t_2}^2, ... ,$ $x^r = x_1^r \ldots x_{t_r}^r $
from $A^*$ and any measure $\theta$ we define $\theta (x^1
\diamond x^2 \diamond \ldots \diamond x^r) = \prod_{i=1}^r
\theta(x^i) .$

 We will use the following well known inequality, whose proof can be found in
\cite{Ga}:

 \emph{ For any two probability distributions $p$ and $q$
 over some alphabet $B$ the following inequality
\begin{equation}\label{cl1}
\sum_{b \in B} p(b) \log \frac{p(b)}{q(b)} \geq 0
\end{equation}
is valid with equality if and only if $p=q.$ }

The value $\sum_{b \in B} p(b) \log \frac{p(b)}{q(b)}$ is often
called Kullback-Leibler divergence.

The following property of the empirical Shannon entropy will be
used later.

\textbf{Lemma}.  \emph{Let $\theta$ be a measure from $M_m(A), m
\geq 0,$ and $x^1, \ldots, x^r$ be words from  $A^*,$ whose
lengths are not less than $m.$  Then
\begin{equation}\label{L}
\theta (x^1 \diamond \ldots \diamond x^r) \leq
 \, 2^{- (t- r m) \, h^*_m(x^1 \diamond
...\diamond x^r)}.\end{equation}}

\subsection{ Codes}
 A data compression method (or code) $\varphi$ is defined as a
set of mappings $\varphi_n $ such that $\varphi_n : A^n
\rightarrow \{ 0,1 \}^*,\, n= 1,2, \ldots\, $ and for each pair of
different words $x,y \in A^n \:$ $\varphi_n(x) \neq \varphi_n(y)
.$  It is also required that each sequence
$\varphi_n(u_1)\varphi_n(u_2) ...\varphi_n(u_r), r \geq 1,$ of
encoded words from the set $A^n, n\geq 1,$ could be uniquely
decoded into $u_1u_2 ...u_r$. Such codes are called uniquely
decodable. For example, let $A=\{a,b\}$, the code $\psi_1(a) = 0,
\psi_1(b) = 00, $  obviously, is not uniquely decodable. It is
well known
 that if  a code $\varphi$ is uniquely decodable
then  the lengths of the codewords satisfy the following
inequality (Kraft inequality): $ \Sigma_{u \in A^n}\: 2^{-
|\varphi_n (u) |} \leq 1\:,$ see, for ex., \cite{Ga}. (Here and
below $|v|$ is the length of $v$, if $v$ is a word and the number
of elements of $v$ if $v$ is a set.) It will be convenient to
reformulate this property as follows:

\emph{Let $\varphi$ be a uniquely decodable code over
an alphabet $A$. Then  for any integer $n$ there exists a measure
$\mu_\varphi$ on $A^n$ such that
\begin{equation}\label{kra}
 - \log \mu_\varphi (u) \:  \leq  \: |\varphi (u)|
\end{equation} for any $u$ from $A^n \,.$}

It is easy to see that  it is true for the measure $\mu_\varphi
(u) = 2^{- |\varphi (u) |} / \Sigma_{u \in A^n}\: 2^{- |\varphi
(u)|} .$
In what follows we call
 uniquely decodable codes just "codes".

We suppose that any code is defined for each sequence of words
$x^1 \diamond x^2 \diamond ... \diamond  x^l.$ (For example, any
code $\varphi$ can be extended to this case
 as follows: $\varphi(x^1 \diamond x^2 \diamond ... \diamond
 x^l)$ $
 = \varphi(x^1)\varphi(x^2) ... \varphi(x^l).$)

  There exist
 so-called universal codes. To introduce these codes we first recall that (as
it is known in Information Theory)  sequences $x_1 \dots x_t,$
generated by a  source $p,$ can be "compressed" up to the
length
$-
\log
 p(x_1 ... x_t)$ bits; on the other hand, for any source $p$ there is no code $\psi$ for
 which the average codeword
 length $  \Sigma
 _{u \in A^t}
 \, p(u) |\psi(u)|$
 is less than  $ - \Sigma_{u \in A^t} \, p(u)\log
 p(u)$.
 Universal codes can reach the lower bound
 $- \log  p(x_1 ... x_t)$ asymptotically for any stationary and ergodic source $p$
 with  probability 1.

  A formal
definition is as follows: A code $\varphi$ is universal if for any
stationary and ergodic source $p$
\begin{equation}\label{un} \lim_{t \rightarrow \infty}\: t^{-1} (- \log
 p(x_1 ... x_t) - |\varphi(x_1 ... x_t)| ) \, = \,0 \end{equation}
 with  probability 1. So, informally speaking, universal
 codes estimate the probability characteristics of the source $p$ and
 use them for efficient "compression".
 One of the first universal
 codes was described in \cite{Ry0}, see also \cite{Ry1}, and now
 there are many efficient universal codes
 and universal
 predictors connected with them,  see \cite{JS,Ki,
 No,Ri,Ry2,Sa}.

\section{Tests For A Finite Alphabet}

\subsection{Goodness-of-fit testing or identity testing}

 Now we consider the problem of testing the hypothesis $H_0^{id}$ 
that the source has a particular distribution $\pi, \pi \in
M_\infty(A),$ against  $H_1^{id}$  that the source is stationary
and ergodic  and differs from $\pi$.
  Let the required  level of significance (or the  Type I error)
   be $\alpha ,\, \alpha \in (0,1).$  We
describe a statistical test which can be constructed based on any
code $\varphi$.

The  main idea of the suggested test is quite natural:
compress a sample  $\bar{x}$ by a code $\varphi$. If the
length of the codeword $|\varphi(\bar{x})|$ is significantly
less than the value $- \log \pi(\bar{x}),$ then $H_0^{id}$
should be rejected. The key observation is that the
probability of all rejected samples is quite small for any
$\varphi$, that is why the Type I error can be made small. The
formal description of the test is as follows:

{ \it Let there be  a sample $\bar{x}$ presented by sequences
$x^1= x^1_1 \ldots x^1_{t_1}, $ $\ldots,$ $x^l= x^l_1 \ldots
x^l_{t_l}, $ generated independently by a source.
 The hypothesis
$H_0^{id}$  is accepted if
\begin{equation}\label{t1} - \log
\pi(\bar{x}) - |\varphi (\bar{x}) | \leq - \log \alpha
.\end{equation}
  Otherwise, $H_0^{id}$ is rejected.}
  We denote this test by $T_{\varphi}^{\,id}(A, \alpha).$

\textbf{Theorem 1.} { \it i) For each distribution  $\pi, \alpha
\in (0,1)$ and  a code $\varphi$, the Type I error of the
described test $T_{\varphi}^{\,id}(A, \alpha)$ is not larger than
$\alpha $ and ii) if, in addition, $\pi$ is a finite-order
stationary and ergodic process over $A^\infty$ (i.e. $\pi \in
M^*(A)$), $\varphi$ is a universal code  then the Type II error of
the test $T_{\varphi}^{\,id}(A, \alpha)$ goes to 0 as the sample
size $\,t \: ( t =\,\sum_{i=1}^l t_i\,)$ tends to infinity.}

\subsection{Testing of serial independence}

Let there be  a sample $\bar{x}$ presented by sequences $x^1=
x^1_1 \ldots x^1_{t_1}, $ $\ldots,$ $x^l= x^l_1 \ldots x^l_{t_l},
$ generated independently by a
 (unknown) source
 and let $t=\sum_{i=1}^l t_i .$ The main hypothesis
$H_0^{SI}$ is that the source  is Markovian, whose order is not
greater than $m,\: (m \geq 0),$ and the alternative hypothesis
$H_1^{SI}$ is that the sample $\bar{x}$ is generated by a
stationary and ergodic source whose order is   greater than $m$
(i.e. the source belongs to $M_\infty(A)  \backslash M_m(A)$). The
suggested test is as follows.

\emph{Let $\varphi$ be any code.
 By definition, the hypothesis $H_0^{SI}$ is accepted if
\begin{equation}\label{cr}
 (t - m \,l)\: h^*_{m}(\bar{x}) -  |\varphi(\bar{x})|  \leq
\log (1 / \alpha) \,,
\end{equation} where $\alpha \in (0,1) .$  Otherwise, $H_0^{SI}$ is rejected.}
 We denote this
test by $T_{\, \varphi}^{\,SI}(A,\alpha).$

\textbf{Theorem 2.} \emph{ i) For any code $\varphi$ the Type I
error of the test $T_{\, \varphi}^{\,SI}(A,\alpha)$ is less than
or equal to $\alpha, \alpha \in (0,1)$ and, ii) if, in addition,
$\varphi$ is a universal code and the sample size $t$ tends to
infinity, then the Type II error goes to 0. }

\subsection{Independence testing}

Now we consider the problem of the  independence testing for
Markovian sources. It is supposed that the source alphabet $A$ is
the Cartesian product of $d$  alphabets $A_1, ..., A_d, $ i.e. $A
= \prod_{i=1}^d A_i,\, d \geq 2\,  $ and it is known a priori that
the source belongs to $M_m(A)$ for some known $m, m \geq 0.$ We
present each letter $x$ as $x = (x^{(1)}, \ldots, x^{(d)}),$ where
$x^{(j)} \in A_j .$ The hypothesis $H_0^{ind}$ is that   $\mu \in
M_m(A)$ is such a source that for each $a = (a^{(1)}, \ldots,
a^{(d)}) \in \prod_{i=1}^d A_i $ and each $x_1 ... x_m \in A^m$
the following equality is valid:
\begin{equation}\label{ind} \mu(x_{m+1} = (a^{(1)}, \ldots, a^{(d)}) |\, x_1 ... x_m)
= \prod_{i=1}^d \mu^{(i)}(x_{m+1}^{(i)} = a^{(i)} |\, x_1 ... x_m)
,\end{equation} where, by definition,
\begin{equation}\label{mu} \mu^{(i)}(x_{m+1}^{(i)} =  a |\, x_1 ... x_m) =\end{equation}
$$ \sum_{b_1,
\ldots, b_{i-1} \in \prod_{j=1}^{i-1} A_j \,} \sum_{\, b_{i+1},
\ldots, b_d \in \prod_{j=i+1}^d A_j}
 \mu(x_{m+1} = (b_1, \ldots, b_{i-1}, a, b_{i+1}, \ldots,
b_d ) |\, x_1 ... x_m)  . $$ The hypothesis $H_1^{ind}$ is that
 the equation (\ref{ind}) is not
valid at least for one $(a^{(1)}, \ldots, a^{(d)}) $ $\in
\prod_{i=1}^d A_i \,$ and $\, x_1 ... x_m \in A^m.$

Let us describe the test  for hypotheses $H_0^{ind}$ and
$H_1^{ind}$. Suppose that there is  a sample $\bar{x}$ presented
as sequences $x^1= x^1_1 \ldots x^1_{t_1}, $ $\ldots,$ $x^l= x^l_1
\ldots x^l_{t_l}, $ generated independently by a source, where, in
turn, any $x_i^j = (x_i^{j(1)}, ...,x_i^{j(d)} ) $. We define $ t
=\,\sum_{i=1}^l t_i\, $ and $\bar{x}^{(k)} = x^{1(k)}_1 \ldots
x^{1(k)}_{t_1}\diamond $ $\ldots \diamond$ $x^{l(k)}_1 \ldots
x^{l(k)}_{t_l}$ for $k = 1, 2, ..., d.$

\emph{Let $\varphi$ be any code.
 By definition, the hypothesis $H_0^{ind}$ is accepted if
\begin{equation}\label{cr2}
\sum_{k=1}^d   (t- m \,l\,) \:h^*_{m}(\bar{x}^{(k)}) -
|\varphi(\bar{x})| \leq \log (1 / \alpha) \,,
\end{equation}  $\alpha \in (0,1) .$
Otherwise, $H_0^{ind}$ is rejected.}
 We denote this
test by $T_{\, \varphi}^{\,ind}(A,\alpha).$ First we give an
informal explanation of the main idea of the test. The Shannon
entropy is the lower bound of the compression ratio and the
empirical entropy $h^*_{m}(\bar{x}^{(k)})$ is its estimate. So, if
$H_0^{ind}$ is true, the sum $\sum_{k=1}^d   (t-m \,l\,)\:
h^*_{m}(\bar{x}^{(k)})$ is, on average, close to the lower bound.
Hence, if the length of a codeword of some code $\varphi$ is
significantly less than the sum of the empirical entropies, it
means that there is some dependence between components, which is
used for some additional compression. The following theorem
describes the properties of the suggested test.

\textbf{Theorem 3.} \emph{i) For any code $\varphi$ the Type I
error of the test $T_{\, \varphi}^{\,ind}(A,\alpha)$ is less than
or equal to $\alpha, \alpha \in (0,1),\,$ and ii) if, in addition,
$\varphi$ is a universal code and $\,t \: $ tends to infinity,
then the Type II error of the test $T_{\,
\varphi}^{\,ind}(A,\alpha)$ goes to 0. }
\subsection{Homogeneity testing}

Let there be $r$ samples $x^1 = x_1^1 \ldots x_{t_1}^1,$ $x^2
= x_1^2 \ldots x_{t_2}^2, ... ,$ $x^r = x_1^r \ldots x_{t_r}^r
, \,( r \geq 2)\,, $ and it is assumed that they are generated by 
 Markovian sources, whose orders  are not larger than $m,\: (m
\geq 0)$ and $m$ is known a priory (i.e. the sources belong to
$M_m(A) $). The null hypothesis $H_0^{hom}$ is that all samples
are generated by one source, whereas the alternative hypothesis
$H_1^{hom}$ is that at least two samples are generated by
different sources.

Let us describe the test  for hypotheses $H_0^{hom}$ and
$H_1^{hom}$. \emph{Let $\varphi$ be any code, $ t =\,\sum_{i=1}^r
t_i\,$  and $\alpha \in (0,1)$.
 By definition, the hypothesis $H_0^{hom}$ is accepted if
\begin{equation}\label{cr3}
  (t \,- \,m \, r)\:
 h^*_{m}(x^1 \diamond x^2  \diamond ... \diamond x^r) -
 \sum_{i=1}^r \,|\varphi(x^i)|\: \leq \,\log (1 / \alpha) \,
.\end{equation} Otherwise, $H_0^{hom}$ is rejected.}
 We denote this
test by $T_{\, \varphi}^{\,hom}(A, \alpha).$

\textbf{Theorem 4.} \emph{i) For  any code $\varphi$ the Type I
error of the test $T_{\, \varphi}^{\,hom}(A, \alpha)$ is less than
or equal to $\alpha, \alpha \in (0,1)$ and ii) if, in addition,
$\varphi$ is a universal code and the sample size $ t $
 goes to infinity in such a way that there
exists a positive constant $c$ for which
\begin{equation}\label{t4} c < t_j / t \end{equation}
 for each $j$, then the Type II error of the test
$T_{\, \varphi}^{\,hom}(A, \alpha)$ goes to 0.}

Let us give some comments concerning the constant $c.$ In fact,
the existence of such a constant  means that all samples are
 present and grow.
 Otherwise, some samples could have a negligible
length, say, 1 letter and, obviously, it would be difficult to
build a reasonable test for such a case.

The suggested test can be extended for a case where it is known
beforehand that some sequences (from $x^1, x^2,  \ldots, x^r $)
were generated by the same source. In this case the same test can
be applied, but the condition ii) can be weaken as follows: for
each source the inequality (\ref{t4}) must be valid for at least
one sample.

\section{Infinite Alphabet}

In this part we consider the case where the source alphabet
$A$ is infinite, say, a part of $R^n$. Our strategy is to use
finite partitions of $A$ and to consider hypothesis
corresponding to the partitions. The main problem of this
approach is as follows: if someone combines letters (or
states) of a Markov chain, the chain order (or memory) can
increase. For example, if an alphabet contains three letters,
there exists  a Markov chain of order one, such that combining
 two letters into one subset transfers the chain into a process
with infinite memory. On the other hand, the main
part of results described above is valid for finite-order
processes. That is why in this part we will consider i.i.d.
processes only (i.e. processes from $M_0(A)$).

In order to avoid numerous repetitions, we will consider a general
scheme, which can be applied to all
  tests using  notations $H_0^\aleph, H_1^\aleph$
 and $T_{\varphi}^{\,\aleph}(A, \alpha),$
 where $\aleph$
is an abbreviation of one of the described tests (i.e. \emph{id,
SI, ind} and \emph{hom}.)

 Let us  give some definitions. Let
$\Lambda = \lambda_1, ..., \lambda_s$ be a finite (measurable)
partition of $A$ and let $\Lambda(x)$ be an element of the
partition $\Lambda,$ which contains $x \in A.$ For any process
$\pi$ we define a process $\pi_\Lambda$ over a new alphabet
$\Lambda$ by equation
$$ \pi_\Lambda( \lambda_{i_1}... \lambda_{i_k}) = \pi( x_1 \in \lambda_{i_1}, ...,
x_k \in \lambda_{i_k} ),$$ where $x_1 ... x_k \in A^k .$ (Such
partitions are widely  used in information theory; see, for
ex., \cite{DV1,DV,Ga} for a  detailed description.)

 We will consider an
infinite sequence  of partitions $\hat{\Lambda}= \Lambda_1,
\Lambda_2, ....$
 and say that such a  sequence discriminates between a
 pair of hypotheses $H_0^\aleph(A), H_1^\aleph(A)$ about processes from
 $M_0(A),$ if for each process $\varrho, $ for which $H_1^\aleph(A)$ is true,
 there exists a partition $\Lambda_j$ for which $H_1^\aleph(\Lambda_j)$
 is true for the process $\varrho_{\Lambda_j}.$ We also define
 a probability distribution $\{\omega =
\omega_1, \omega_2, ... \}$ on integers $\{ 1, 2, ... \}$ by
\begin{equation}\label{om} \omega_1 = 1 - 1/ \log 3,\: ... \,,\:
\omega_i\,= 1/ \log (i+1) - 1/ \log (i+ 2),\: ... \; .
\end{equation}
 (In what follows we will use this distribution, but
the theorem described below is obviously true for   any
distribution with nonzero probabilities.)

Let $H_0^\aleph(A), H_1^\aleph(A)$ be a pair of hypotheses,
$\hat{\Lambda}= \Lambda_1, \Lambda_2, ...$ be a sequence of
partitions, $\alpha $ be from $ (0,1)$ and $\varphi$ be a code.
The scheme for all the tests is as follows:

 { \it
 The hypothesis
$H_0^{\aleph}(A)$  is accepted if for all $i = 1, 2, 3, ... $  the
test $T_\varphi^\aleph (\Lambda_i, (\alpha \omega_i) )$ accepts
 the hypothesis $H_0^\aleph(\Lambda_i).$
  Otherwise, $H_0^\aleph$ is rejected.} We denote this test as $\textbf{T}_{
\alpha,\varphi}^{\aleph}(\hat{\Lambda}).$

\emph{Comment.} It is important to note that one does not need to
check an infinite number of inequalities when one 
 applies
this test. The point is that the hypothesis $H_0^{\aleph}(A)$
has to be accepted if the left part in (\ref{t1}), (\ref{cr}),
(\ref{cr2}) and (\ref{cr3}), correspondingly,  is less than $-
\log (\alpha
\omega_i).$ Obviously, $- \log (\alpha \omega_i)$ goes to infinity
if $i$ increases. That is why there are many cases, where   it
is enough to check a finite number of hypotheses
$H_0^{\aleph}(\Lambda_i)$.

\textbf{Theorem 5.} { \it i) For each  $\alpha \in (0,1),$
sequence of partitions $\hat{\Lambda}$ and  a code $\varphi$, the
Type I error of the described test $\textbf{T}_{
\alpha,\varphi}^{\aleph}(\hat{\Lambda})$ is not larger than
$\alpha $ and ii) if, in addition, $\varphi$ is a universal code
and $\hat{\Lambda}$ discriminates between $H_0^\aleph(A),
H_1(A)^\aleph,$ then the Type II error of the test $\textbf{T}_{
\alpha,\varphi}^{\aleph}(\hat{\Lambda})$ goes to 0, when the
sample size tends to infinity (in the case of the homogeneity
testing, in addition,  the inequality (\ref{t4}) should be valid).
}

\section{The Experiments }

In this part we describe results of some experiments and a
simulation study carried out to estimate an efficiency of the
suggested tests. The obtained results show that the described
tests as well as the suggested approach in general can be used in
applications.

\subsection{Randomness testing}

First we consider the problem of randomness testing, which is
a particular case of goodness-of-fit testing. Namely, we will
consider a null hypothesis $H^{rt}_0$ that a given bit
sequence is generated by Bernoulli source with equal
probabilities of 0 and 1 and the alternative hypothesis
$H^{rt}_1$ that the sequence is generated by a stationary and
ergodic source which differs from the source under $H^{rt}_0$.
This problem
 is important for  random number (RNG) and pseudorandom
number generators (PRNG) testing and there are many methods for
randomness testing suggested in  literature.
 Thus,
recently National Institute of Standards and Technology (NIST,
USA) suggested "A statistical test suite for random and
pseudorandom number generators for cryptographic
applications", see \cite{rng}.

We investigated
 linear
congruent generators (LCG), which are defined by the following
equality $$X_{n+1}=(A*X_n+C)\: mod\, M ,$$ where $X_{n}$ is the
$n$-th generated number \cite{K}. Each such  generator we will
denote by $LCG(M,A,C,X_0),$  where $X_0$ is the initial value of
the generator. Such generators are well studied and many of them
are used in practice, see  \cite{Kn}.

In our experiments we extract an eight-bit word from each
generated $X_i$ using the following algorithm. Firstly, the number
$\mu = \lfloor M/256 \rfloor $ was calculated and then each $X_i$
was transformed into an 8-bit word $\hat{X}_i$ as follows:

\begin{equation}\label{tr} \left.
\begin{array}{cc}
\hat{X}_i = \lfloor X_i/256 \rfloor\,\; if X_i < 256 \mu
\\
\hat{X}_i = empty\; word \,\; if X_i \geq 256 \mu
\end{array}
\right\} \end{equation}
 Then a sequence was compressed by the archiver \emph{ACE  v
1.2b} (see http://www.winace.com/). Experimental data about
testing of three linear congruent generators is given in the table
1.

\begin{table}[h]

\caption{ Pseudorandom number generators testing }
\begin{tabular}{|c|c|c|}
\hline
$\;\;\qquad\qquad$ parameters $\qquad\qquad$ / length (bits) &400
000& 8 000 000
\\ M,A,C, $ X_0 $ & $ \quad $  & $ \quad $
\\ $ 10^8 +1,23,0,47594118 \qquad\quad$ & 390 240    & 7635936
\\ $ 2^{31},2^{16}+3,0,1 \quad \quad$ & extended & 7797984
\\ $ 2^{32},134775813,1,0 \quad \quad$ & extended & extended
\\
\hline
\end{tabular}
\end{table}

So, we can see from the first line of the table that the
$400000-$bit sequence generated by the LCG($ 10^8 +1,23,0,47594118
$) and transformed according to (\ref{tr}), was compressed to a
$390 240-$bit sequence. (Here  400000 is the length of the
sequence after transformation.) If we take the level of
significance $ \alpha \geq 2^{- 9760} $ and apply the test
$T_{\varphi}^{\,id}( \{0,1\}, \alpha),$ ($\varphi = \emph{ACE  v
1.2b}$),  the hypothesis $H^{rt}_0$ should be rejected, see
Theorem 1 and (\ref{t1}). Analogously, the second line of the
table shows that the $8000000-$bit sequence generated by LCG($
2^{31},2^{16}+3,0,1 $) cannot be considered  random
($H^{rt}_0$ should be rejected if the level of significance
$\alpha$ is greater than $2^{- 202016} $). On the other hand,
the suggested test accepts $H^{rt}_0$ for the sequences
generated by the third generator, because the lengths of the
``compressed'' sequences increased.

The obtained information corresponds to the known data about the
considered generators. Thus, it is shown in \cite{Kn} that the
first two generators are bad whereas the third generator was
investigated in \cite{mo}  and is regarded as good. So, we can see
that the suggested testing is quite efficient.

In a recently published paper \cite{RM} the described method
was applied for testing random number and pseudorandom number
generators and its efficiency was compared with the mentioned
methods from "A statistical test suite for random and
pseudorandom number generators for cryptographic applications"
\cite{rng}. The point is that the tests from \cite{rng} are
selected basing on comprehensive theoretical and experimental
analysis and can be considered as the state-of-the-art in
randomness testing. It turned out that the suggested tests,
which were based on archivers RAR and ARJ, were more powerful
than many methods recommended by NIST in \cite{rng}; see
\cite{RM} for details.

\subsection{Simulation study of  serial independence testing }

A selection of the simulation results concerning independence
tests is presented in this part.
 We generated binary sequences by the first order Markov
source with different probabilities (see table 2 below) and
applied the test $T_{\, \varphi}^{\,SI}(\{ 0,1\},\alpha)$ to
test 
 the hypothesis $H^{SI}_0$ that a given bit sequence is
generated by Bernoulli source  and the alternative hypothesis
$H^{SI}_1$ that the sequence is generated by a stationary and
ergodic source which differs from the source under $H^{SI}_0$.

We tried several different archivers and the universal code $R $
described in Appendix 2. It turned out that the power of the code
$R$ is larger than the power of the tried archivers, that is why
we present results for the test $T_{\, R}^{\,SI}(\{
0,1\},\alpha),$ which is based on this code, for $\alpha = 0.01.$
The table 2 contains results of calculations.

\begin{table}[h]
\caption{Serial independence testing for Markov source of order 6
 ("rej" means rejected, "acc" - accepted. In all cases $p(x_{i+1}=0 | x_{i}=1)= 0.5 $ )}
\begin{tabular}{|c|c|c|c|c|c|}
\hline
$\;\;\qquad\qquad$ probabilities $\qquad\qquad$ / length (bits) &
$2^{9}$ & $2^{14}$ & $2^{16}$ & $2^{18}$ & $2^{23}$
\\ $ p(x_{i+1}=0 | x_{i}=0)= 0.8\quad  $ &
rej & rej & rej& rej & rej\\
  $ p(x_{i+1}=0 | x_{i}=0)=
0.6 \quad $ &
acc & rej & rej & rej & rej \\
$ p(x_{i+1}=0 | x_{i}=0)= 0.55 \;$ &
acc & acc & rej & rej & rej \\
$ p(x_{i+1}=0 | x_{i}=0)= 0.525\:$ &
acc & acc & acc & rej & rej \\
$ p(x_{i+1}=0 | x_{i}=0)= 0.505\:$ &
acc & acc & acc & acc & rej \\
 \hline
\end{tabular}
\end{table}

We know that the source is Markovian and, hence, the hypothesis
$H^{SI}_0$ (that a  sequence is generated by Bernoulli source) is
not true.
 The table shows how  the value of the Type II error
depends on the sample size and the source  probabilities.

The similar calculations were carried out for the Markov source of
order 6. We applied the test $T_{\, \varphi}^{\,SI}(\{
0,1\},\alpha), \:\alpha = 0.01,\,$  for checking the hypothesis
$H^{SI}_0$ that a given bit sequence is generated by Markov source
of order at most 5 and the alternative hypothesis $H^{SI}_1$ that
the sequence is generated by a stationary and ergodic source which
differs from the source under $H^{SI}_0$. Again, we know that
$H^{SI}_0$  is not true and the table 3 shows how  the value of
the Type II error depends on the sample size and the source
probabilities.

\begin{table}[h]
\caption{Serial independence testing for Markov source of order 6.
In all cases $p(x_{i+1}=0 \,|\,(\sum_{j=i-6}^i x_{i})\:
\emph{mod}\, 2 \,=1\,)\:= 0.5$. }
\begin{tabular}{|c|c|c|c|c|c|}
\hline
 probabilities $\qquad\qquad \quad
\qquad\qquad\qquad$/ length (bits) & $2^{14}$ & $2^{18}$ &
$2^{20}$ & $2^{23}$ & $2^{28}$
\\ $ p(x_{i+1}=0 |\,(\sum_{j=i-6}^i x_{j})\:
\emph{mod}\, 2 \, = 0) = 0.8 \quad \qquad\qquad\qquad\,$ &
rej & rej & rej& rej & rej\\
  $ p(x_{i+1}=0 | \,(\sum_{j=i-6}^i x_{j})\:
\emph{mod}\, 2 \, = 0) = 0.6  \quad \qquad\qquad\qquad\,$ &
acc & rej & rej & rej & rej \\
$ p(x_{i+1}=0 | \,(\sum_{j=i-6}^i x_{j})\: \emph{mod}\, 2 \, = 0)
= 0.55 \;  \qquad\qquad\qquad\;$&
acc & acc & rej & rej & rej \\
$ p(x_{i+1}=0 |  \,(\sum_{j=i-6}^i x_{j})\: \emph{mod}\, 2 \, =
0)= 0.525\:\qquad\qquad\qquad$ &
acc & acc & acc & rej & rej \\
$ p(x_{i+1}=0 |  \,(\sum_{j=i-6}^i x_{j})\: \emph{mod}\, 2 \, =
0)= 0.505\:\qquad\qquad\qquad$ &
acc & acc & acc & acc & rej \\
 \hline
\end{tabular}
\end{table}


\section{Conclusion.}

In this part we point out some generalizations of the suggested
approach as well as clarify  the connection with some statistical
methods.

Having taken into account the Kraft inequality (\ref{kra}), we can
rewrite the goodness-of-feet test (\ref{t1}) as follows:

\begin{equation}\label{ml}if\;\,
\pi(\bar{x}) / \mu_\varphi(\bar{x}) \geq \, \alpha \,\; then \quad
H_0,\; otherwise \quad H_1, \end{equation}
 where, as before, $\mu_\varphi (\bar{x}) = 2^{-
|\varphi (\bar{x}) |} / \Sigma_{u \in A^t}\: 2^{- |\varphi (u)|},
\: t $ is the sample size.  Clearly, (\ref{ml}) looks like the
likelihood ratio test, which is one of the main statistical tools.
Moreover, all other tests can be presented in the same manner.
Thus, if we denote $2^{ - (t- \, l \,m)\: h^*_{m}(\bar{x})}$ from
(\ref{cr}) by $\pi, $ we can rewrite the serial independence test
(\ref{cr})  in the same form as (\ref{ml}). The same is true for
the independence testing (\ref{cr2}) and homogeneity testing
(\ref{cr3}), if we denote by $\pi$ the values $2^{ - \sum_{k=1}^d
(t- \,l \,m) h^*_{m}(\bar{x}^{(k)} )} $ and $2^{ -  (t \: - \,m \,
r)\,
 h^*_{m}(x^1 \diamond x^2  \diamond ... \diamond x^r)} $,
 correspondingly.

Now we use the representation (\ref{ml}) in order to extend the
suggested tests to the following more general case. Let there be
several codes (or archivers) $\varphi_1, \varphi_2, \ldots,
\varphi_l$ and we want to build a test, which is based on all of
them. In order to get such a test, we define the "mixture"
probability distribution and the mixture distribution of codeword
lengths  by equalities
$$ \mu_{mix}(\bar{x}) = ( 2^{- |\varphi_1(\bar{x})|} +
2^{-|\varphi_2(\bar{x})|} + \ldots +2^{-|\varphi_l(\bar{x})| }
)\,/\,l, \; \quad | \varphi_{mix}(\bar{x})| = - \log
\mu_{mix}(\bar{x}) ,
$$
correspondingly. Obviously, the Kraft inequality (\ref{kra}) is
valid for $|\varphi_{mix}|$ and, therefore, $|\varphi_{mix}|$  can
be used in all suggested tests instead of $|\varphi|.$ In the case
when the set of codes $\varphi_1, \varphi_2, \ldots$ is infinite,
we can use some probability distribution $\tau$ on the set $1,2,3,
...$ and define
$ \mu_{mix}$ and $|\varphi_{mix}|$ as follows:
\begin{equation}\label{mix}
\mu_{mix}(\bar{x}) = \sum_{i=1}^\infty \tau_i \, 2^{-
|\varphi_i(\bar{x})|},   \quad | \varphi_{mix}(\bar{x})| = - \log
\mu_{mix}(\bar{x}). \end{equation} (For example, the distribution
$\omega$ (\ref{om}) can be used here as the distribution $\tau$.)

It can be easily seen from the descriptions of the tests that
their power is grater, if the length of the codeword
$|\varphi(\bar{x})|$ is less. That is why it is natural to
look for a code $\varphi_i$ whose length is minimal. First of
all we can find such a code $\varphi_\delta$ that
\begin{equation}\label{min}
- \log \: (\tau_\delta \:2^{- |\varphi_\delta(\bar{x})|} ) =
\min_i \;( - \log \: (\tau_i \:2^{- |\varphi_i(\bar{x})|} )).
\end{equation}

Having taken into account (\ref{mix}), we can see that
$$- \log \: (\tau_\delta \:2^{-
|\varphi_\delta(\bar{x})|} ) \leq  | \varphi_{mix}(\bar{x})| .$$
If we denote by $\varphi_{mm}$
 the code, whose codeword length $ | \varphi_{mm}(\bar{x})|
 = $ $ - \log \: (\tau_\delta \:2^{-
|\varphi_\delta(\bar{x})|} )$ for each $\bar{x}$, the later
inequality shows that $ | \varphi_{mm}(\bar{x})| \leq $ $|
\varphi_{mix}(\bar{x})| $ for any sample $\bar{x}, $ and, hence,
the power of the tests based on the code $\varphi_{mm}$ is not
less than the power of the tests based on the code
$\varphi_{mix}.$

It is worth noting that  codes $\varphi_{mix}$ and $\varphi_{mm}$
(and corresponding distributions, which  based on the Kraft
inequality (\ref{kra})),  were applied for constructing optimal
universal codes and predictors in \cite{Ry0,Ry1} and later both
constructions were used in  mathematical statistics and related
fields under different names (aggregating strategy, weighted
majority algorithms, etc.).

 One of the reason of a  popularity
 of both constructions is their asymptotical optimality.
 Thus, in case of hypothesis testing, the codes $\varphi_{mix}$ and $\varphi_{mm}$
 give, in a
certain sense,  the most powerful (asymptotically) tests. Indeed,
if we suppose that the family of codes $\varphi_1, \varphi_2, ...$
contains  a code $\varphi_{opt},$ whose codeword length
($|\varphi_{opt}(\bar{x})|$) is minimal (say, with probability
1,when the sample size increases), we can see from the definitions
$\varphi_{mix}$ and $\varphi_{mm}$ that $|\varphi_{mix}(\bar{x})|
\leq |\varphi_{opt}(\bar{x})| + \,\emph{const}$ and
$|\varphi_{mm}(\bar{x})| \leq
|\varphi_{opt}(\bar{x})|+\,\emph{const},$ where $\emph{const}
\:=\:- \,\log \tau_{opt}.$ On the other hand, for any processes
(whose entropy is larger than zero), the codeword length
$|\varphi_{opt}(\bar{x})|$ goes to infinity, if the sample size
($|\, \bar{x}|\,$) increases and, hence, the impact of
\emph{const} decreases.


\section{Appendix 1.  Predictors and Universal Codes}

Let a source generate a message $x_1\ldots x_{t-1}x_t\ldots $,
$x_i\in A$ for all $i$. After the
 first $t$ letters $x_1,\ldots, x_{t-1},x_t$ have been processed
the following letter $ x_{t+1}$ needs to be predicted. By
definition, the prediction is the set of non-negative numbers
$\gamma(a_1|x_1\cdots x_t),\cdots, $ $\gamma(a_n|x_1\cdots x_t)$
which are estimates of the unknown conditional probabilities
$p(a_1|x_1\cdots x_t),$ $ \cdots,$ $p(a_n|x_1\cdots x_t)$, i.e. of
the probabilities $p(x_{t+1}=a_i| x_1\cdots x_t)$; $i=1,\cdots,n$.

Laplace suggested the following predictor:
\begin{equation}\label{L}
L_0(a|x_1\cdots x_t) = (\nu_{x_1\cdots x_t}(a) +1 )/ (t+ |A | ),
\end{equation}  see \cite{FE}. (We use $L_0$ here in order to show that it is
intended to predict sources from $M_0(A)$. Later this predictor
will be extended to $ M_i(A), \: i > 0.$) For example, if $ A=
\{0, 1 \}, \: x_1 ... x_5 = 01010,$ then the Laplace prediction is
as follows: $L_0(x_{6}=0| 01010) = (3+1)/ (5+2) = 4/7, L_0(x_{6}=1
| 01010) = (2+1)/ (5+2) = 3/7.$

 It is natural to  estimate the error of
prediction  by the the Kullback-Leibler (K-L) divergence
 between a distribution $p$ and its estimation. Consider
a source $p$ and a predictor $\gamma$. The {\it{ error }}  is
characterized by the divergence
\begin{equation}\label{r0}
\rho_{\gamma,p}(x_1\cdots x_t)= \sum_{a\in A}p(a|x_1\cdots
x_t)\log\frac{p(a|x_1\cdots x_t)} {\gamma(a|x_1\cdots x_t)}.
\end{equation}
 As we mentioned above,  for
any distributions  $p$ and $\gamma$ the K-L divergence is
nonnegative and equals 0 if and only if $p(x) = \gamma(x)$ for all
$x.$
 For fixed $t$, $r_{\gamma,p}$ is a random variable, because
$x_1, x_2, \cdots, x_t$ are random variables. We define the
average error
  at time $t$ by
\begin{equation}\label{r1}
\rho^t(p\|\gamma)=E\,\left(r_{\gamma,p}(\cdot)\right)=\nonumber \,
\sum_{x_1\cdots x_t\in A^t}p(x_1\cdots
x_t)\,\,\rho_{\gamma,p}(x_1\cdots x_t).
\end{equation}
It is shown in \cite{Ry2} that the error of Laplace predictor goes
to 0 for any i.i.d. source $p$. More precisely, it is proven that
\begin{equation}\label{rL} r^t(p\|L_0)
< (|A| - 1)/ (t + 1) \end{equation}
 for any source $p;\,$
 ( see also \cite{RT}).

For any predictor $\gamma$ we define the corresponding probability
measure by
\begin{equation}\label{me}
\gamma(x_1 ... x_t) = \prod_{i=1}^t \gamma(x_i |\, x_1\cdots
x_{i-1}).
\end{equation} For example, the Laplace measure $L_0$  of the word
$x_1 \ldots x_t = 0101  $ is as follows: $L_0(0101) = \frac{1}{2}
\frac{1}{3} \frac{1}{2} \frac{2}{5} = \frac{1}{30}.$ By analogy
with (\ref{r0}) and (\ref{r1}) we define
\begin{equation}\label{R0} \rho_{\gamma, p}(x_1 ... x_t) =
t^{-1}\: ( \log ( p (x_1 ... x_t) / \gamma (x_1 ... x_t) )
\end{equation} and
\begin{equation}\label{RR} \bar{\rho}_t (\gamma, p) = t^{-1} \sum_{x_1 ... x_t \in A^t} p(x_1 ... x_t)
\log ( p (x_1 ... x_t) / \gamma (x_1 ... x_t) ).
\end{equation}
For example, from those definitions and (\ref{rL}) we obtain the
following estimation for Laplace predictor $L_0$ and any i.i.d.
source $p \,$:
\begin{equation}\label{Lr}
\bar{\rho}_t (L_0, p) < ( \log t  + c)/t,
\end{equation} where $c$ is a constant.

The average error  (\ref{RR}) has three interesting
characteristics. Firstly, it can be easily seen from (\ref{r0}),
(\ref{r1}) and (\ref{RR}) that $\bar{\rho}_t (\gamma, p)$ is the
average error
of the predictor $\gamma$ when it is applied to the process
$p:$
$$ \bar{\rho}_t (\gamma, p) = t^{-1} \sum_{j=1}^t
\rho^j(p\|\gamma).
$$ Secondly, having taken into account the definition of the Shannon
entropy  (\ref{moe}), we can easily see that for $p \in M_0(A)$
\begin{equation}\label{iid} \bar{\rho}_t (\gamma, p) = t^{-1} E_p(- \log \gamma(x_1 ... x_t)) \:
- \: h_0(p). \end{equation} The third characteristic is  connected
with the theory of universal coding. One can construct a  code
with codelength $\gamma_{code}(a|x_1\cdots x_t)\approx -\log_2
\gamma(a|x_1\cdots x_n)$ for any letter $a\in A$ (since Shannon's
original research, it has been well known, cf. e.g. \cite{Ga},
that, using block codes with large block length or more modern
methods of arithmetic coding, the approximation may be as
accurate as you like). If one knows the real distribution $p,$
one can base coding on the true distribution $p$ and not on
the prediction
$\gamma$. The difference in performance measured by average code
length is given by
$$
\sum_{a\in A}p(a|x_1\cdots x_t)(-\log_2 \gamma (a|x_1\cdots x_t))
-\sum_{a\in A}p(a|x_1\cdots x_t)(-\log_2 p(a|x_1\cdots x_t))$$
$$
=\sum_{a\in A}p(a|x_1\cdots x_t)\log_2 \frac{p(a|x_1\cdots x_t)}
{\gamma (a|x_1\cdots x_t)}.$$

Thus this excess,
 it is exactly the error (\ref{r0}) defined above. Analogously,
 if we encode the sequence $x_1 \ldots x_t
$ based on a predictor $\gamma$ the redundancy per letter is
defined by (\ref{R0}) and (\ref{RR}). So, from mathematical point
of view the universal prediction and universal coding are
identical. But
 $- \, \log \gamma (x_1 ... x_t) $ and $- \,\log p
(x_1 ... x_t) $ have a very natural interpretation. The first
value is a code word length (in bits), if the "code" $\gamma$
is applied for compressing of the word $x_1 ... x_t$ and the
second one is the minimally possible codeword length. The
difference is the redundancy of the code  and, at the same
time, the error of the predictor. It is worth noting that
there are many other deep interrelations between  universal
coding, prediction and estimation, see \cite{Ri,Ry1}.

As we saw in (\ref{Lr}), the average error  of the Laplace
predictor is upper bounded by $(|A| - 1) ( \log t + O(1) ) /(t+1),
$ when $t$ grows. Krichevsky suggested the predictor
$K_0(a|x_1\cdots x_t) = (\nu_{x_1\cdots x_t}(a) +1/2 )/ (t+ |A |/2
)$ and showed that the error of this predictor is
asymptotically less: 
$\bar{\rho}_t ({K_0}, p) $ is
upper bounded by
$(|A| - 1) ( \log t + O(1) ) /(2t). $ Moreover, he showed that
this predictor is asymptotically optimal  in the sense that for
any other predictor $\gamma$ there exists a source $\hat{p}$ for
which the error $\bar{\rho}_t (\gamma, \hat{p}) $ is not less than
$(|A| - 1) ( \log t + O(1) ) /(2t),$ see \cite{Kr}.

From definitions (\ref{L}) and (\ref{me}) we can see that the
Laplace predictor ascribes the following probabilities:
\begin{equation}\label{Lp}
L_0(x_1 ... x_t) = \prod_{i=1}^{t} \frac{ \nu_{x_1 ... x_{i-1}}
(x_i)+ 1 }{i-1+|A|}= \frac{\prod_{a \in A} (\nu_{x_1 ... x_t}(a)
)!}{((t+ |A|-1)!)/ (|A| - 1)!} \,.
\end{equation}
Analogously, for $K_0$ we obtain
\begin{equation}\label{Kp}
K_0(x_1 ... x_t) =\prod_{i=1}^{t} \frac{ \nu_{x_1 ... x_{i-1}}
(x_i)+ 1/2 }{i-1+|A|/2} = \frac{\prod_{a \in A} (\prod_{j=
1}^{\nu_{x_1 ... x_t}(a)} (j- 1/2))}{\prod_{i= 0}^{t-1}(i+ |A|/2)}
\,.
\end{equation}
The following simple example shows the difference between the
predictors: If $A= \{ 0,1 \}$ and $x_1 \ldots x_t = 0101 , $ then
$L_0$ and $K_0$  ascribe the
 probabilities $ \frac{1}{2} \frac{1}{3} \frac{1}{2} \frac{2}{5} = \frac{1}{30}$
 and $ \frac{1}{2} \frac{1}{4} \frac{1}{2} \frac{3}{8} =  \frac{3}{128},$
 correspondingly.

The product $ (r+1/2) ((r + 1) + 1/2) ... (s - 1/2) $ can be
presented as a ratio $\frac{\Gamma(s+1/2)}{\Gamma(r+1/2)},$ where
$\Gamma(\:)$ is the gamma function (see for definition, for ex.,
\cite{Kn} ). So, (\ref{Kp}) can be presented as follows:
 \begin{equation}\label{Kp1}
K_0(x_1 ... x_t) = \frac{\Gamma(|A|/2)}{\Gamma(1/2)^{|A|}}\:
\frac{\prod_{a \in A} \Gamma(\nu_{x_1 ... x_t}(a)+ 1/2
)}{\Gamma((t+ |A|/2 )} \,.
\end{equation}
As we mentioned above the average error of the Krichevsky
predictor is asymptotically minimal. That is why we will focus our
attention on this predictor and, for the sake of completeness, we
prove an upper bound for its error.

\textbf{Claim 1}.  For any stationary and ergodic  source
generating letters from a finite alphabet $A$ the average
error
 of $K_0$  is upper bounded as
follows: $$ -\: t^{-1} \sum_{x_1 ... x_t \in A^t} p(x_1 ... x_t)
\log ( K_0(x_1 ... x_t)) - h_0(p) \leq ((|A| - 1) \log t +C )/ (2
t) ,
$$ where $C$ is a constant.

\textit{Proof} is given in the Appendix 2.

\textit{Comment}.  In particular, if the source is i.i.d., the
average error is less than $(( |A| - 1) \log t +C )/ (2 t); $ see
(\ref{iid}).

We indicated that extensions of both predictors to cover the
general Markov case are possible.
We take this up now.
  The trick
is to view a Markov source $p\in M_m(A)$ as resulting from $|A|^m$
i.i.d. sources. We illustrate this idea by an example from
\cite{RT}. So assume that $A=\{O,I\}$, $m=2$ and assume that the
source $p\in M_2(A)$ has generated the sequence
$$ OOIOIIOOIIIOIO. $$
We represent this sequence by the following four subsequences:
$$ **I*****I***** ,$$
$$ ***O*I***I***O ,$$
$$ ****I**O****I* ,$$
$$ ******O***IO** .$$
These four subsequences contain letters which follow
$OO$,  $OI$,  $IO$ and  $II$, respectively.
 By definition,
$p\in M_m(A)$ if $p(a |x_1\cdots x_t) = p(a|x_{t-m+1}\cdots x_t
$), for all $0 < m \leq t $, all $a\in A$ and all $x_1\cdots x_t
 \in A^t $. Therefore, each of the four generated subsequences
may be considered to be generated by a Bernoulli source. Further,
it is possible to reconstruct the original sequence if we know the
four ($=|A|^m$) subsequences and the two ($=m$) first letters of
the original sequence.

Any predictor $\gamma $ for i.i.d. sources can be applied for
Markov sources. Indeed, in order to predict, it is enough to store
in the memory $|A|^m$ sequences, one corresponding to each word in
$A^m$. Thus, in the example, the letter $x_3$ which follows
$OO$ is predicted based on the Bernoulli method $\gamma$
corresponding to the $x_1 x_2$- subsequence ($=OO$),
 then $x_4$ is predicted based on the Bernoulli
method corresponding to $x_2x_3$, i.e. to the $OI$-
subsequence, and so forth. When this scheme is applied along
with either  $L_0$ or  $K_0$ we denote the obtained predictors
as
$L_m$ and $K_m,$ correspondingly and define the probabilities
for the first $m$ letters as follows: $ L_m(x_1) = L_m(x_2) =
\ldots L_m (x_m) = 1/|A|\,, $ $K_m(x_1) = K_m(x_2) = \ldots
K_m(x_m) = 1/|A|\,. $

 Having taken into account
(\ref{Lp}) and (\ref{Kp1}), we can present the Laplace and
Krichevsky predictors for $M_m(A)$ as follows:
\begin{equation}\label{lm}
L_m(x_1 ... x_t) =\cases{\frac{1}{|A|^t},&if $t \leq m\,$;\cr
 & \cr
           \frac{1}{|A|^m} \prod_{v \in A^m} \frac{\prod_{a \in A}
           (\nu_x(v a ))!}{((\bar{\nu}_x(v  )+|A|)!)/ (|A|-1)!}, &if $ t > m $ \, ,}
\end{equation}
\begin{equation}\label{km}
K_m(x_1 ... x_t) =\cases{\frac{1}{|A|^t},&if $t \leq m\,$;\cr
 & \cr
 \frac{1}{|A|^m}\: ( \frac{\Gamma(|A|/2)}{\Gamma(1/2)^{|A|}}
)^{|A|^m} \prod_{v \in A^m} \frac{\prod_{a \in A}\:
           ( \Gamma( \nu_x(v a )+ 1/2)}{( \Gamma( \bar{\nu}_x(v  )+|A|/2))}, &if $ t > m $ \, ,}
\end{equation}
where $\bar{\nu}_x(v  )= \sum_{a \in A} \nu_x(v a ), \:x = x_1 ...
x_t. $

We have seen that any source from $M_m(A)$ can be presented as
a "sum" of $|A|^m$ an i.i.d. sources. From this we can easily
see that the error of a predictor for the source from $M_m(A)$
can be upper bounded by the error of i.i.d. source multiplied
by $|A|^m$. In particular, we obtain from Claim 1 the
following upper bound.

 \textbf{Claim 2}.   For any stationary and ergodic  source
generated letters from a finite alphabet $A$  the average error
 of the Krichevsky  predictor $K_m$  is upper bounded as
follows: $$ -\:  t^{-1} \sum_{x_1 ... x_t \in A^t} p(x_1 ... x_t)
\log ( K_m (x_1 ... x_t)) - h_m(p) \leq  |A|^m ((|A| - 1) \log t +
C)/ (2 t) ,
$$ where $C$ is a constant.

Now we can describe the universal predictor $R$ and code
$R_{code}$ from \cite{Ry0,Ry1}. By definition,
 $$ R(x_1 ... x_t) = \sum_{i=0}^\infty \, \omega_{i+1} \:K_i(x_1 ...
 x_t) , $$
 $$R(x_t |\, x_1 ... x_{t-1}) = R(x_1 ... x_t) / R(x_1 ... x_{t-1})
 $$ and $ | R_{code}(x_1 ... x_t) | = - \log R(x_1 ... x_t).
 $ It is worth
 noting that this construction can be applied to the Laplace
 predictor
 (if we use $L_i$ instead of $K_i$) and any other family of predictors (or
 codes).

 \textbf{Claim 3.}
  \emph{Let  the predictor $R$ be applied to a source $p $.
  Then, for  any stationary and ergodic source $p \in M_\infty (A)$
 the
 error (\ref{RR}) of the predictor $R$ goes to 0, when the sample
 size $t $ goes to $ \infty$.
}


 \emph{Proof } can be derived from  Claim 2 and the properties
 of the Shannon entropy. Indeed, we can see from the definition of
 $R$ and  Claim 2 that the average error
  is upper bounded as
follows: $$ - \;  t^{-1} \sum_{x_1 ... x_t \in A^t} p(x_1 ... x_t)
\log ( R (x_1 ... x_t)) - h_k(p)$$ $$ \leq  (|A|^k (|A| - 1) \log
t + \log (1/ \omega_i) + C)/ (2 t) ,
$$ for any $k= 0, 1, 2, ...$. Taking into account that for any $p \in M_\infty(A)$
$\lim_{k\rightarrow\infty}h_k(p) = h_\infty(p),$ we can see that
$$ (\lim_{t\rightarrow\infty}t^{-1} \sum_{x_1 ... x_t \in A^t} p(x_1 ... x_t)
\log ( R (x_1 ... x_t)) - h_\infty(p)) = 0.$$ The main property of
the universal codes  (\ref{un}) is also true for $R_{code}$
and can be easily  derived from  Claim 3  using   standard
techniques of  ergodic theory.

\section{Appendix 2.    Proofs}

\textit{Proof} of the Lemma. First we show that for any source
$\theta^* \in M_0(A)$ and any words $x^1 = x^1_1 ... x^1_{t_1},$
$... ,$ $x^r = x^r_1 ... x^r_{t_r},$
$$\theta^* (x^1 \diamond ... \diamond x^r) = \prod_{a \in A}
(\theta^*(a))^{\nu_{x^1 \diamond ... \diamond x^r}(a)} $$
\begin{equation}\label{ta} \leq \prod_{a \in A} (
\nu_{x^1 \diamond ... \diamond x^r}(a)/t)^{\nu_{x^1 \diamond ...
\diamond x^r}(a)} ,
\end{equation} where $t = \sum_{i=1}^r t_i .$
Here the equality holds, because $\theta^* \in M_0(A)$ . The
inequality follows from  (\ref{cl1}). Indeed, if $ p(a) = \nu_{x^1
\diamond ... \diamond x^r}(a)/t$ and $ q(a) = \theta^* (a),$ then
$$ \sum_{a \in A} \frac{\nu_{x^1 \diamond ... \diamond x^r}(a)}{t}
\log \frac{(\nu_{x^1 \diamond ... \diamond x^r}(a)/t)}{\theta^*(a)
}\geq 0. $$ From the latter inequality we obtain (\ref{ta}).
Taking into account the definition (\ref{He1}) and (\ref{ta}), we
can see that the statement of Lemma is true for this particular
case.

For any $\theta \in M_m(A)$ and $x = x_1 \ldots x_s, \,s > m,$ we
present $\theta (x_1 \ldots x_s)$ as $\theta (x_1 \ldots x_s)=
\theta(x_1 \ldots x_m) \prod_{u \in A^m } \prod_{a \in A} \theta
(a|\,u)^{\nu_x(ua)}\:,$ where $\theta(x_1 \ldots x_m)$
 is
the limit
 probability of the word $x_1 \ldots x_m .$ Hence,
$\theta (x_1 \ldots x_s) \leq \prod_{u \in A^m } \prod_{a \in A}
\theta (a|\,u)^{\nu_x(ua)}\:.$ Taking into account the inequality
(\ref{ta}), we obtain $\prod_{a \in A} \theta (a|\,u)^{\nu_x(ua)}
\leq \prod_{a \in A} ( \nu_x(ua)/\bar{\nu}_x(u))^{\nu_x(ua)}$
 for any word $u$. Hence, $$ \theta (x_1 \ldots x_s) \leq \prod_{u \in A^m }
 \prod_{a \in A} \theta (a|\,u)^{\nu_x(ua)} $$ $$
\leq  \prod_{u \in A^m } \prod_{a \in A} (
\nu_x(ua)/\bar{\nu}_x(u))^{\nu_x(ua)}.$$
If we apply those inequalities to $\theta(x^1 \diamond ...
\diamond x^r),$ we immediately obtain the following
inequalities
$$ \theta(x^1 \diamond ... \diamond x^r)  \leq \prod_{u \in A^m }
 \prod_{a \in A} \theta (a|\,u)^{\nu_{x^1 \diamond ... \diamond x^r}(ua)}
\leq $$ $$ \prod_{u \in A^m } \prod_{a \in A} ( \nu_{x^1 \diamond
... \diamond x^r}(ua)/\bar{\nu}_{x^1 \diamond ... \diamond
x^r}(u))^{\nu_{x^1 \diamond ... \diamond x^r}(ua)}.$$ Now the
statement of the Lemma follows from the definition  (\ref{He1}).

\textit{Proof} of Theorem 1.  In order to avoid cumbersome
notations we first consider a case where the sample $\bar{x}$
is one sequence $x_1 ... x_t$ and then note how the proof can
be extended for the general case. Let $C_\alpha$ be a critical
set of the test $T_{\varphi}^{\,id}(A, \alpha)$, i.e., by
definition, $ C_\alpha = \{ u: u \in A^t \,\, \:\& \: - \log
\pi(u) - |\varphi (u) | > - \log \alpha  \}. $ Let
$\mu_\varphi$ be a measure for which (\ref{kra}) is true.
We define an auxiliary set $
\hat{C}_\alpha  $ $ = \{ u:  - \log \pi(u) - (- \log
\mu_\varphi(u) ) $ $ > - \log \alpha  \}. $ We have $ 1 \geq $ $
\sum_{u \in \hat{C}_\alpha} \mu_\varphi(u) $ $ \geq \sum_{u \in
\hat{C}_\alpha} \pi(u) / \alpha $ $ = (1/ \alpha)
\pi(\hat{C}_\alpha) .$ (Here the second inequality follows from
the definition of $\hat{C}_\alpha,$ whereas all others are
obvious.) So, we obtain that $\pi(\hat{C}_\alpha) \leq \alpha .$
From  definitions of $C_\alpha, \hat{C}_\alpha $ and (\ref{kra})
we immediately obtain that $\hat{C}_\alpha \supset C_\alpha.$
Thus, $\pi(C_\alpha) \leq \alpha .$ By definition, $\pi(C_\alpha)$
is the value of the Type I error. The first statement of the
theorem 1 is proven.

Let us prove the second statement of the theorem. Suppose that the
hypothesis $H_1^{id}(A)$ is true. That is, the sequence $x_1
\ldots x_t$ is generated by some stationary and ergodic source
$\tau$ and $\tau \neq \pi .$ Our strategy is to show that
\begin{equation}\label{inf}    \lim _{t\rightarrow\infty}- \log \pi(x_1 \ldots
x_t) - |\varphi(x_1 \ldots x_t)| = \infty
\end{equation}
 with probability 1 (according to the
measure $\tau$). First we represent (\ref{inf}) as
$$
 - \log \pi(x_1
\ldots x_t) - |\varphi(x_1 \ldots x_t)| $$ $$  = t ( \frac{1}{t}
\log \frac{\tau(x_1 \ldots x_t)}{\pi(x_1 \ldots x_t)}   +
\frac{1}{t}( - \log \tau(x_1 \ldots x_t)  - |\varphi(x_1 \ldots
x_t)| ) ). $$
 From
this equality and the property of a universal code (\ref{un}) we
obtain
\begin{equation}\label{inf2}  - \log \pi(x_1 \ldots
x_t) - |\varphi(x_1 \ldots x_t)| = t\, ( \frac{1}{t} \log
\frac{\tau(x_1 \ldots x_t)}{\pi(x_1 \ldots x_t)}  + o(1)).
\end{equation}
 From  (\ref{moe}) -- (\ref{smb}) we can see that
\begin{equation}\label{inf3} \lim_{t\rightarrow\infty} -  \log \tau(x_1 \ldots x_t)
/t   \leq h_k( \tau)
\end{equation} for any $k \geq 0$ (with probability 1).
It is supposed that the process $\pi$ has a finite memory, i.e.
belongs to $M_s(A)$ for some $s$. Having taken into account the
definition of $M_s(A)$ (\ref{ma}), we obtain the following
representation: $$ - \log \pi(x_1 \ldots x_t)/t  =   - t^{-1}
\sum_{i=1}^t \log \pi(x_i|\,x_1 \ldots x_{i-1}) $$ $$  =  - t^{-1}
(\sum_{i=1}^k \log \pi(x_i|\,x_1 \ldots x_{i-1})  + \sum_{i=k+1}^t
\log \pi(x_i|\,x_{i-k} \ldots x_{i-1})) $$ for any $k \geq s.$
According to the ergodic theorem there exists a limit $$
\lim_{t\rightarrow\infty} t^{-1} \sum_{i=k+1}^t \log
\pi(x_i|\,x_{i-k} \ldots x_{i-1}),$$
 which  is equal to $ h_k(\tau),$ see
\cite{Billingsley, Ga}. So, from the two latter equalities we can
see that $$  \lim_{t\rightarrow\infty} (- \log \pi(x_1 \ldots
x_t))/t  = - \sum_{v \in A^k} \tau(v) \sum_{a \in A} \tau(a|\,v)
\log \pi(a|\,v).$$ Taking into account this equality, (\ref{inf3})
and (\ref{inf2}), we can see that $$- \log \pi(x_1 \ldots x_t) -
|\varphi(x_1 \ldots x_t)|   \geq  t \,(\sum_{v \in A^k} \tau(v)
\sum_{a \in A} \tau(a|\,v) \log (\tau(a|\,v) / \pi(a|\,v) )) +
o(t)
$$ for any $k \geq s.$
   From this
inequality and (\ref{cl1}) we can obtain that  $\:- \log \pi(x_1
\ldots x_t) - |\varphi(x_1 \ldots x_t)| \geq c\: t + o(t)$, where
$c$ is a positive constant, $t\rightarrow\infty.$ Hence,
(\ref{inf}) is true.

Let us consider a case where $ \bar{x}$ is a sequence
$ x^1 = x^1_1 ... x^1_{t_1},$ $... ,$ $x^l = x^l_1 ... x^l_{t_l}$
(i.e. $\bar{x}= x^1\diamond \ldots \diamond x^l). $ The proof of
the first statement of the theorem is analogical and can be simply
repeated for this case. In order to prove the second
statement we note that the length of at least one sequence
$x^i$ goes to infinity and, hence, the equality (\ref{inf}) is
true for that sequence, whereas for all other sequences the
differences
$\log \pi(x^j) - |\varphi(x^j)|$ are either bounded or go to
infinity. The theorem is proven.

\textit{Proof} of Theorem 2.

 We only consider a case
where the sample $\bar{x}$ is one sequence $x_1 ... x_t,$ because
the general case is analogical, but requires  cumbersome
notations. Let us denote the critical set of the test $T_{\,
\varphi}^{\,SI}(A,\alpha)$  as $C_\alpha,$ i.e., by definition, $
C_\alpha = \{ x_1 \ldots x_t :\; (t - m)\: h^*_m(x_1 \ldots x_t) -
|\varphi(x_1 ... x_t)| )
>
\log (1 / \alpha)   \} . $ From  (\ref{kra}) we can see that there
exists such a measure $\mu_\varphi $ that $ - \log \mu_\varphi(x_1
... x_t)$ $ \leq |\varphi(x_1 ... x_t)| \,. $ We also define
\begin{equation}\label{C} \hat{C}_\alpha = \{ x_1 \ldots x_t :\;
(t - m)\: h^*_m(x_1 \ldots x_t) - (- \log \mu_\varphi(x_1 ...
x_t)) \,)
  >
\log (1 / \alpha)   \} . \end{equation} Obviously, $\hat{C}_\alpha
\supset C_\alpha .$ Let $\theta$ be any source from $M_m(A).$ The
following chain of equalities and inequalities is true: $$ 1 \geq
\mu_\varphi(\hat{C}_\alpha) = \sum_{x_1 \ldots x_t \in
\hat{C}_\alpha} \mu_\varphi(x_1 \ldots x_t) $$ $$ \geq \alpha^{-1}
\sum_{x_1 \ldots x_t \in \hat{C}_\alpha} 2^{(t-m) h_m^*(x_1 \ldots
x_t)} \geq \alpha^{-1} \sum_{x_1 \ldots x_t \in \hat{C}_\alpha}
\theta(x_1 \ldots x_t) = \theta(\hat{C}_\alpha).$$
  (Here both equalities and the first inequality are
 obvious, the second   and the third inequalities follow from
(\ref{C}) and the Lemma, correspondingly.) So, we obtain that
$\theta(\hat{C}_\alpha) \leq \alpha $ for any source $\theta \in
M_m(A).$ Taking into account that $\hat{C}_\alpha \supset
C_\alpha,$ where $C_\alpha$  is the critical set of the test, we
can see that the probability of the  Type I error is not greater
than $\alpha.$ The first statement of the theorem is proven.

The proof of the second  statement  will  be based on some results
of Information Theory. We obtain from (\ref{un}) and (\ref{smb})
that for any stationary and ergodic $p$
\begin{equation}\label{unh}
\lim_{t\rightarrow\infty}  t^{-1} |\varphi(x_1 ... x_t)| =
h_\infty (p)
\end{equation} with probability 1.
    It can be seen from (\ref{He}) that $h^*_m$
 is an estimate for the $m-$order Shannon
entropy  (\ref{moe}). Applying the ergodic theorem we obtain $
\:\lim_{t\rightarrow\infty} h^*_m (x_1 \ldots x_t ) = h_m(p)$ with
probability 1; see \cite{Billingsley, Ga}. It is known in
Information Theory that $h_m(\varrho) - h_\infty(\varrho)
> 0, $ if $\varrho$ belongs to $M_\infty(A)\: \backslash \:M_m(A),$
see \cite{Billingsley, Ga}. It is supposed that $H^{SI}_1(A)$ is
true, i.e. the considered process belongs to $M_\infty(A)\:
\backslash\: M_m(A).$ So, from (\ref{unh}) and  the last equality
we obtain that $\lim_{t\rightarrow\infty} ( (t - m) \,h^*_m (x_1
\ldots x_t )  - |\varphi(x_1 ... x_t)|) = \infty .$ This proves
the second statement of the theorem.

\textit{Proof} of Theorem 3. As before, we only consider a case
where the sample $\bar{x}$ is one sequence $x_1 ... x_t,$ because
the general case is analogical. Let $C_\alpha$ be a critical set
of the test, i.e., by definition, $C_\alpha = \{ (x_1,..., x_t): $
 $
\sum_{i=1}^d $ $ (t-m) h^*_{m}(x_1^{(i)} ... x_t^{(i)}) $ $ -
|\varphi(x_1 ... x_t)| > \log (1 / \alpha)  \} .$ There exists a
measure $\mu_\varphi,$  for which (\ref{kra}) is valid. Hence,
\begin{equation}\label{CC} C_\alpha \subset C_\alpha^* \equiv \{
(x_1,..., x_t):  \sum_{i=1}^d (t-m) h^*_{m}(x_1^{(i)} ...
x_t^{(i)}) - \log (1/ \mu_\varphi(x_1,..., x_t) > \log (1 /
\alpha) \}.
\end{equation} Let $\theta$ be any measure from $M_m(A).$
Then
 $$ 1
\geq \mu_\varphi (C_\alpha^*)  \geq \alpha^{-1} \sum_{x_1,..., x_t
\in C_\alpha^*} \prod_{i=1}^d 2^{- (t-m) h^*_{m}(x_1^{(i)} ...
x_t^{(i)}) }.$$ Having taken into account the Lemma, we obtain $$
1 \geq \mu_\varphi (C_\alpha^*) \geq \sum_{x_1,..., x_t \in
C_\alpha^*} \prod_{i=1}^d \theta^i(x_1^{(i)} ... x_t^{(i)}) \:.$$
It is supposed that $H^{ind}_0$  is true and, hence, (\ref{ind})
is valid. So, from the latter inequalities we can see that $ 1
\geq \mu_\varphi (C_\alpha^*) \geq  \sum_{x_1,..., x_t \in
C_\alpha^*} \theta(x_1,..., x_t) .$ Taking into account that
$\sum_{x_1,..., x_t \in C_\alpha^*} \theta(x_1,..., x_t) =
\theta(C_\alpha^*) $ and (\ref{CC}), we obtain that
$\theta(C_\alpha) \leq \alpha. $ So, the first statement of the
theorem is proven.

We give a short scheme of the proof of the second statement of the
theorem, because it is based on well-known facts of Information
Theory. It is known that $h_m(\mu) - \sum_{i = 1}^d h_m(\mu^i) = 0
$   if $H_0^{ind}$ is true and this difference is negative under
$H_1^{ind}.$ A universal code compresses a sequence till $ t
h_m(\mu)$ (Informally, it uses dependence for the better
compression.) That is why the difference $( \sum_{i = 1}^d
h_m(\mu^i)\,- \,t \:h_m(\mu) )$ goes to infinity, when $t$
increases and, hence, $H_0^{ind} $ will be rejected.

\textit{Proof} of Theorem 4. For short, we consider a case of two
samples and i.i.d. sources (i.e. $m= 0$), because a generalization
is obvious. So, there are two samples $x^1 = x^1_1 ... x^1_{t_1}$
and $x^2 = x^2_1 ... x^2_{t_2}$  generated by sources from
$M_0(A).$ As before, let $C_\alpha$ be a critical set of the test,
i.e., by definition, $C_\alpha = \{ (x^1, x^2): $ $ (t_1 + t_2) \:
$ $ h_0(x^1 \diamond x^2) - (|\varphi(x^1)| + |\varphi(x^2)|) $ $
> \log (1/\alpha) \}.$ There
exists a measure $\mu_\varphi$ for which (\ref{kra}) is valid. So,
$C_\alpha $ $ \supset C_\alpha^*  $ $ \equiv \{ (x^1, x^2): $ $
(t_1 + t_2) \: $ $ h_0^*(x^1 \diamond x^2) -
(\log(1/\mu_\varphi(x^1)) + \log(1/\mu_\varphi(x^2)) ) $ $ > \log
(1/\alpha) \}.$ Let us suppose that $H_0^{hom}$ is true. It means
that $(x^1, x^2)$ are created by some source $\theta \in M_0(A).$
Having taken into account the definition of the set $C_\alpha^* $
and Lemma, we obtain the following chain of inequalities:
$$ 1 \geq \mu_\varphi (C_\alpha^*) = \sum_{(x^1,x^2) \in C_\alpha^*}
\mu_\varphi (x^1 \diamond x^2)  \geq $$ $$\alpha^{-1}
\sum_{(x^1,x^2) \in C_\alpha^*} 2^{- (t_1 + t_2) h_0^*(x^1
\diamond x^2)} \geq   \sum_{(x^1,x^2) \in C_\alpha^*} \theta(x^1
\diamond x^2) = \theta(C_\alpha^*).$$ Hence, $\theta(C_\alpha^*)
\leq \alpha$ and, taking into account that the critical set
$C_\alpha \subset C_\alpha^*, $ we finish the proof of the first
statement of the theorem.

Let us suppose that $H_1^{hom}$ is true, i.e. the samples $x^1,
x^2$ are generated by different sources $\theta_1, \theta_2$,
correspondingly. For any $\gamma \in (0,1)$ we define
$\theta_\gamma = \gamma \theta_1 + (1-\gamma) \theta_2$ and let
\begin{equation}\label{del}  \delta = \inf_{\gamma \in [c, 1-c]} (\: h_0(\theta_\gamma) -
(h_0(\theta_1) + h_0(\theta_2))\:)\,, \end{equation} where $c$ is
defined in (\ref{t4}). Due to the Jensen  inequality for the
Shannon entropy, we can easily see that $\delta > 0.$ Having taken
into account the definition of a universal code and ergodicity of
$\theta_1, \theta_2$ we obtain that $$ (t_1 + t_2) h_0^*(x^1
\diamond x^2) - (|\varphi(x^1)| + |\varphi(x^2)| ) =  (t_1 +
t_2)\,( h_0(\frac{t_1}{t_1 + t_2} \theta_1 + \frac{t_2}{t_1 + t_2}
\theta_2) \:\; - $$ $$ ( \frac{t_1}{t_1 + t_2} h_0(\theta_1) +
\frac{t_2}{t_1 + t_2} h_0(\theta_2)\, ) \:)+ o(t_1 + t_2),
$$  (with probability 1), if $(t_1 + t_2) \rightarrow \infty$. Taking into account the definition
(\ref{del}) and (\ref{t4}) we obtain from the last equality that
$$
 (t_1 + t_2)\, h_0^*(x^1
\diamond x^2) - (|\varphi(x^1)| + |\varphi(x^2)| )
 \,
>
\delta \: (t_1 + t_2)+o(t_1 + t_2). $$ Hence, the difference $$
 (t_1 + t_2) h_0^*(x^1
\diamond x^2) - (|\varphi(x^1)| + |\varphi(x^2)| ) $$ goes to
infinity and  the second statement of the theorem is proven.

\textit{Proof} of Theorem 5. The following chain proves the first
statement of the theorem:
$$ Pr \{ H_0^{\aleph}(A)\quad is\; rejected \;/ H_0 \:is\: true\} = $$ $$ Pr
\{\bigcup_{i=1}^\infty  \{H_0^\aleph(\Lambda_i)\quad is\;
rejected\;/ H_0 \:is\: true\} \:\}\, \leq $$
$$ \sum_{i=1}^\infty Pr
 \{ H_0^\aleph(\Lambda_i)\;/ H_0 \:is\: true\} \, \leq \sum_{i=1}^\infty (\alpha \omega_i) \, =
\alpha .
$$ (Here both inequalities follow from the description of the
test, whereas the last equality follows from (\ref{om}).)

The second statement  also follows from the description of the
test. Indeed, let a sample be created by a source
$\varrho,$ for which $ H_1(A)^\aleph$ is true. It is supposed
that the sequence of partitions $\hat{\Lambda}$ discriminates
between
$H_0^\aleph(A), H_1^\aleph(A).$ By definition, it means that there
exists $j$ for which $H_1^\aleph(\Lambda_j)$
 is true for the process $\varrho_{\Lambda_j}.$ It  immediately
 follows from Theorem 1-4
  that the Type II error
 of the test $T_{\,
\varphi}^{\,\aleph}(\Lambda_j,\alpha \omega_j)$ goes to 0, when
the sample size tends to infinity.

\textit{Proof} of Claim 1. From (\ref{Kp1}) we obtain:
$$ - \log
K_0(x_1 ... x_t) = - \log (
\frac{\Gamma(|A|/2)}{\Gamma(1/2)^{|A|}}\: \frac{\prod_{a \in A}
\Gamma( \nu_{x_1 ... x_t}(a) + 1/2 )}{\Gamma((t+ |A|/2 )} )
$$
$$ = c_1 + c_2 |A| + \log \Gamma (t+ |A|/2) - \sum_{a \in A}
\Gamma(\nu_{x_1 ... x_t}(a) + 1/2) ,
$$ where $c_1 , c_2$ are constants.  Now we use the well known Stirling formula
$$ \ln \Gamma(s) = \ln \sqrt{2 \pi} + (s- 1/2)\ln s   - s +
\theta/ 12, $$ where $\theta  \in (0,1),$ see, for ex., \cite{Kn}.
Using this formula we rewrite the previous equality as
$$ - \log
K_0 (x_1 ... x_t) = - \sum_{a \in A} \nu_{x_1 ... x_t}(a) \log
(\nu_{x_1 ... x_t}(a)/t) + (|A|-1)\log t /2 + \bar{c}_1 +
\bar{c}_2 |A| ,$$ where $\bar{c}_1 , \bar{c}_2$ are constants.
Having taken into account the definition of the empirical entropy
(\ref{He}), we obtain
$$ - \log
K_0(x_1 ... x_t) \leq t h^*_0( x_1 \ldots x_t) + (|A|-1)\log t /2
+ c |A| .$$ Hence,
$$ \sum_{x_1 \ldots x_t \in A^t}p(x_1 \ldots x_t)( - \log (K_0 (x_1 ...
x_t) )) $$ $$ \leq t ( \sum_{x_1 \ldots x_t \in A^t}p(x_1 \ldots
x_t) h^*_0( x_1 \ldots x_t) + (|A|-1)\log t /2 +  c |A| .$$ Having
taken into account the definition (\ref{He}), we apply the well
known Jensen inequality for the concave function $ - x \log x $
and obtain the following inequality:
$$ \sum_{x_1 \ldots x_t \in A^t}p(x_1 \ldots x_t)( - \log (K_0 (x_1 ...
x_t) ) \leq  $$ $$   - t (\sum_{x_1 \ldots x_t \in A^t}p(x_1
\ldots x_t) ((\nu_{x_1 ... x_t}(a) / t)) \log \sum_{x_1 \ldots x_t
\in A^t}p(x_1 \ldots x_t) (\nu_{x_1 ... x_t}(a) / t) + (|A|-1)\log
t /2 +  c |A| .
$$ The source $p$ is stationary and ergodic, so the average
frequency $\sum_{x_1 \ldots x_t \in A^t}p(x_1 \ldots x_t) \nu_{x_1
... x_t}(a)$ is equal to $p(a)$ for any $a \in A $ and we obtain
from two last formulas the following inequality:
$$ \sum_{x_1 \ldots x_t \in A^t}p(x_1 \ldots x_t)( - \log (K_0(x_1 ...
x_t) )  \leq t h_0(p) + (|A|-1)\log t /2 +  c |A| $$ ( where
$h_0(p) = - \sum_{a \in A} p(a) \log p(a) $ is the
 Shannon entropy). Claim 1 is proven.

\section{Acknowledgment } The authors wish to thank  Andrey Gruzin
and Viktor Monarev who carried out all experiments described in
the part 5.


\begin{thebibliography}{1}

\bibitem{Ba}
G. J. Babu, A. Boyarsky, Y. P. Chaubey, P. Gora,  New statistical
method for filtering and entropy estimation of a chaotic map from
noisy data, International Journal of Bifurcation and Chaos
14(11)(2004) 3989-3994.

\bibitem{Billingsley} P. Billingsley,  Ergodic theory and information,
 John Wiley \& Sons, 1965.

\bibitem{V1}
 R. Cilibrasi, P.M.B. Vitanyi,
 Clustering by Compression, IEEE Transactions on Information Theory
 51(4)(2005) 1523-1545.

\bibitem{V2}
R. Cilibrasi, R. de Wolf,  P.M.B. Vitanyi,  Algorithmic Clustering
of Music,
   Computer Music Journal 28(4) (2004) 49-67.



\bibitem{CS} I. Csisz$\acute{a}$r, P. Shields,
The consistency of the BIC Markov order estimation, Annals of
Statistics, 6(2000)  1601-1619.


\bibitem{DV1} G.A. Darbellay, I. Vajda,  Entropy expressions
for multivariate continuous distributions, Research Report no.
1920, UTIA, Academy of Science, Prague, 1998.
(library@utia.cas.cz).

\bibitem{DV} G.A. Darbellay, I. Vajda, Estimatin of the
mutual information with data-dependent partitions, IEEE Trans.
Inform. Theory 48(5)(1999) 1061-1081.


\bibitem{e}  M. Effros, K. Visweswariah, S.R. Kulkarni, S. Verdu,
 Universal lossless source coding with the Burrows Wheeler
transform, IEEE Trans. Inform. Theory 48(5) (2002) 1061 - 1081.


\bibitem{FE}
W. Feller,  An Introduction to Probabability Theory and Its
Applications, vol.1., John Wiley \& Sons, New York, 1970.



\bibitem{Fi}
B.M. Fitingof, Optimal encoding for unknown and changing
statistica of messages, Problems of Information Transmission
2(2)(1966) 3-11.

\bibitem{Ga}
R.G. Gallager, Information Theory and Reliable Communication, John
Wiley \& Sons, New York, 1968.

\bibitem{GKR}
 K. Ghoudi,  R.J. Kulperger, B. Remillard,   A Nonparametric Test of
Serial Independence for Time Series and Residuals, Journal of
Multivariate Analysis 79(2)(2001) 191-218.

\bibitem{JS}
 P. Jacquet, W. Szpankowski,  L. Apostol, Universal predictor based on
pattern matching, IEEE Trans. Inform. Theory 48(2002) 1462-1472.


\bibitem{KS}
 M.G. Kendall, A. Stuart,  The advanced theory of statistics;
Vol.2: Inference and relationship, London, 1961.

\bibitem{Ki}
J. Kieffer,   Prediction and Information Theory, Preprint, 1998.
(available at
ftp://oz.ee.umn.edu/users/kieffer/papers/prediction.pdf/ )

\bibitem{K-Y}
 J.C. Kieffer, En-Hui Yang,  Grammar-based codes: a new class of
universal lossless source codes, IEEE Transactions on Information
Theory, 46(3)(2000) 737 - 754.

\bibitem{Kn}
 D.E. Knuth,   The art of computer programming, Vol.2,
Addison Wesley, 1981.


\bibitem{K}
A.N. Kolmogorov, Three approaches to the quantitative definition
of information, Problems Inform. Transmission 1(1965) 3-11.

\bibitem{Kr}
R. Krichevsky, Universal Compression and Retrival, Kluver Academic
Publishers, 1993.

\bibitem{Ku}
S. Kullback, Information Theory and Statistics, Wiley, New York,
1959.


\bibitem{Ma}
U.  Maurer, Information-Theoretic Cryptography, in: Advances in
Cryptology - CRYPTO '99, Lecture Notes in Computer Science,
Springer-Verlag,  1666(1999),  47-64.



 \bibitem{mo}
O. Moeschlin, E.  Grycko, C.  Pohl, F. Steinert, Experimental
Stochastics, Springer-Verlag, Berlin Heidelberg, 1998.


\bibitem{Mo}
 G. Morvai, S.J. Yakowitz,  P.H. Algoet,   Weakly convergent
nonparametric forecasting of stationary time
     series. IEEE Trans. Inform. Theory 43(1997) 483 -
     498.

\bibitem{No}
 A.B. Nobel,   On optimal sequential prediction, IEEE Trans.
Inform. Theory 49(1)(2003)  83-98.

\bibitem{Ri}
J. Rissanen,   Universal coding, information, prediction, and
estimation,  IEEE Trans. Inform. Theory 30(4)(1984) 629-636.

\bibitem{Ri2}
J.Rissanen, Hypothesis selection and testing by the MDL principle,
The Computer Journal, 42(4) (1999) 260-269.

\bibitem{rng}
 A. Rukhin, J. Soto, J. Nechvatal, M. Smid,  D. Banks,
A statistical test suite for random and pseudorandom number
generators for cryptographic applications, NIST Special
Publication 800-22, 2001.
 http://csrc.nist.gov/rng/SP800-22b.pdf

\bibitem{Ry0}
 B.Ya. Ryabko,  Twice-universal coding, Problems of
Information Transmission 20(3)(1984) 173-177.



\bibitem{Ry1}
B.Ya. Ryabko,  Prediction of random sequences and universal
coding, Problems of Inform. Transmission 24(2)(1988) 87-96.

\bibitem{Ry2}
B.Ya. Ryabko,  A fast adaptive coding algorithm, Problems of
Inform. Transmission, 26(4)(1990) 305-317.

\bibitem{RA}
 B. Ryabko, J. Astola, Universal Codes as a Basis for
Nonparametric Testing of Serial Independence for Time Series,
Journal of Statistical Planning and Inference.(Submitted)

\bibitem{RM}
B. Ryabko, V. Monarev,  Using Information Theory Approach to
Randomness  Testing, Journal of Statistical Planning and
Inference, 133(1)(2005)   95-110.

\bibitem{RR}
B. Ryabko,  Zh. Reznikova, Using Shannon Entropy and Kolmogorov
Complexity  To Study the Communicative System and Cognitive
Capacities in  Ants,  Complexity 2(2)(1996) 37-42.




\bibitem{RT}
 B. Ryabko, F. Topsoe,  On Asymptotically Optimal Methods of
Prediction and Adaptive Coding for Markov Sources, Journal of
Complexity 18(1)(2002) 224-241.

\bibitem{Sa}
 S.A. Savari, A probabilistic approach to some
asymptotics in noiseless communication, IEEE Transactions on
Information Theory 46(4)(2000) 1246-1262.


\bibitem{S}
 C. E. Shannon, A mathematical theory of communication,  Bell Sys. Tech.
J., 27(1948) 379-423,623-656.

\bibitem{S2}
C.E. Shannon, Communication theory of secrecy systems, Bell Sys.
Tech. J. 28(1948) 656-715.

\bibitem{Shi}
P.C. Shields, The interactions between ergodic theory and
information theory, IEEE Transactions on Information Theory
44(6)(1998)  2079 - 2093.




\end{thebibliography}
\end{document}